\documentclass[prd,twocolumn,english,superscriptaddress,floatfix]{revtex4-1}

\usepackage{amsmath,amssymb,mathrsfs}
\usepackage{amsfonts,bm}
\usepackage{amsmath}
\usepackage{amssymb}
\usepackage{amsthm}
\usepackage{graphicx}
\usepackage{graphics}
\usepackage{subfigure}
\usepackage{color}
\usepackage{bm}
\usepackage{hyperref}
\usepackage{rotating}
\usepackage{comment}
\usepackage[colorinlistoftodos]{todonotes}
\usepackage{ tipa }
\usepackage{ upgreek }

\usepackage[normalem]{ulem}

\usepackage{xfrac}

\newcommand{\be}{\begin{equation} }
\newcommand{\ee}{\end{equation} }
\newcommand{\ba}{\begin{eqnarray} }
\newcommand{\ea}{\end{eqnarray} }

\newcommand{\bpm}{\begin{pmatrix}}
\newcommand{\epm}{\end{pmatrix}}
\newcommand{\bmm}{\begin{matrix}}
\newcommand{\emm}{\end{matrix}}

\newcommand{\la}{\label}
\newcommand{\p}{\partial}

\newcommand{\bea}{\begin{eqnarray}}
\newcommand{\eea}{\end{eqnarray}}

\makeatother

\usepackage{babel}

\begin{document}

%%%%%%%%%%%%%%%%%%%%%%%%%%%%%%%%%
\title{Topological fluids with boundaries and fractional quantum Hall edge dynamics: A fluid dynamics derivation of the chiral boson action}
%%%%%%%%%%%%%%%%%%%%%%%%%%%%%%%%%

\author{Gustavo M. Monteiro}
 \affiliation{Department of Physics, City College, City University of New York, New York, NY 10031, USA }

\author{V. P. Nair}
 \affiliation{Department of Physics, City College, City University of New York, New York, NY 10031, USA }
 \affiliation{CUNY Graduate Center, New York, NY 10031}
 
  \author{Sriram Ganeshan} 
 \affiliation{Department of Physics, City College, City University of New York, New York, NY 10031, USA }
\affiliation{CUNY Graduate Center, New York, NY 10031}

\date{\today}

%%%%%%%%%%%%%
\begin{abstract}

This paper investigates the bulk and boundary dynamics of Laughlin states, which are modeled using composite boson theory within a fluid dynamics framework. In this work, we adopt an alternative starting point based on a hydrodynamic action with topological terms, which fleshes out the fluid aspects of the Laughlin state manifestly. For a particular choice of the velocity field, the fluid equation for this action is akin to first-order hydrodynamic equations, supplemented with an additional constitutive equation known as the Hall constraint. When a hard wall boundary is present, one of the topological terms in the fluid action triggers anomaly inflow, indicating the presence of gauge anomaly at the edge. The first-order hydrodynamic equations require a second boundary condition which, in the absence of dissipation, can be either a no-slip or a no-stress condition. We find that the no-slip condition, where the fluid adheres to the wall is incompatible with the chiral edge dynamics. On the other hand, the no-stress condition, which allows the fluid to move along the wall without friction, is consistent with the expected chiral edge dynamics of the Laughlin state. %This results in the gauge anomaly generating tangential forces on the boundary that are offset by fluid stresses. 
Furthermore, our work derives this modified no-stress boundary condition within a variational principle. This is accomplished by incorporating a chiral boson action within the boundary action that is non-linearly coupled to the edge density, thus systematically extending the edge chiral Luttinger liquid theory.

\end{abstract}
%%%%%%%%%%%%%
\maketitle

%%%%%%%%%%%%%%%%%%%%%%%%
%%%%%%%%%%%%%%%%%%%%%%%%

\section{Introduction} 

Under specific conditions, the ground state of an interacting many-body system can exhibit fluid dynamic behavior. These quantum fluids, commonly called superfluids, are characterized by their dissipationless nature, meaning they flow without experiencing shear or bulk viscosities. The most well-known example of a superfluid is Helium-II~\cite{helium-book}. However, phenomena such as superconductivity~\cite{abrikosov1957magnetic} and the fractional quantum Hall (FQH) effect~\cite{stone1990superfluid} can also be understood as manifestations of quantum fluid behavior. 

A specific class of FQH states, known as Laughlin states, is frequently characterized in terms of composite bosons. The condensate dynamics of these bosons can be effectively described by hydrodynamic equations with an additional constitutive relation called the Hall constraint~\cite{stone1990superfluid, abanov2013effective}. This constraint ties the superfluid vorticity to fluctuations in the condensate density. These superfluid equations can also be derived from the Chern-Simons-Ginzburg-Landau (CSGL) theory. This theory describes FQH states with filling fractions $\nu=\frac{1}{2k+1}$, where $k$ is a positive integer. In the CSGL model, these states are described by composite bosons coupled to a Chern-Simons gauge field, a procedure known as flux attachment~\cite{zhang1989effective, read1989order, zhang1992chern}. The composite boson approach complements the composite fermion approach, wherein the Laughlin state is interpreted as a collective state of composite particles comprising an electron with two attached flux quanta. While both the composite boson and composite fermion approaches are believed to produce similar qualitative results for the Laughlin states, the composite fermions picture has no superfluid or conventional hydrodynamic interpretation~\cite{jain2007composite}. We will focus on this composite boson framework from here on in. 

One of us recently showed that the same fluid dynamic equations can also be derived from a hydrodynamic action containing topological terms. This variational principle is expressed in auxiliary fields known as Clebsch potentials~\cite{nair2021topological}.  We consider this hydrodynamic variational principle as an alternative starting point for studying the universal physics of the composite Boson model of the Laughlin states. Specifically, we investigate how the gauge anomaly manifests at the edge when enforcing hydrodynamic boundary conditions. By exploring the implications of these boundary conditions, we aim to gain insights into the fundamental nature of FQH states and their associated phenomena. 

Before we discuss boundary conditions, it is important to point out that, in this framework, the FQH fluid velocity is not defined {\it a priori}, leading to distinct (yet qualitatively equivalent) forms of the fluid dynamical equations depending on the chosen velocity parameterizations. In this work, we choose a particular form of superfluid velocity that ensures the resulting stress tensor is, first-order in gradient expansion, commonly referred to as first-order hydrodynamics. This choice of velocity is specifically engineered to cancel out the second-order derivative terms in the stress tensor known as the quantum pressure (Madelung) terms~\cite{geracie2014effective, monteiro2023hamiltonian}. Furthermore, this choice allows a more convenient examination of the different boundary conditions.  

In ordinary first-order hydrodynamics, when fluids are confined within rigid walls, the usual boundary conditions consist of the no-penetration condition, which requires the normal component of the velocity to vanish at the boundary, along with either the no-slip or the no-stress boundary condition. The no-slip condition implies that the fluid sticks to the wall, meaning the tangential velocity must be zero at the boundary. On the other hand, the no-stress condition allows the fluid to slip at the wall, as long as the flow does not generate tangential forces at the boundary. Both the no-slip and no-stress boundary conditions do not do any work, and they can be derived systematically from a variational principle. Typically, for classical fluids, determining whether to use the no-stress or no-slip condition depends on the specific details of the fluid interface. However, in this work, we use the boundary gauge anomaly as the criterion to derive the appropriate boundary conditions for the FQH fluid. 

In the presence of an external electric field, the hydrodynamic action contains additional topological terms that modify the conventional no-penetration condition, resulting in the anomaly inflow mechanism. In this mechanism, a tangent electric field induces a normal current into the boundary, causing charge to accumulate at the edge, which is known as the gauge anomaly. This accumulated charge is subsequently forced to flow along the edge due to the same tangent electric field, which directly contradicts the no-slip condition. 

On the other hand, the no-stress condition takes into account the existence of density fluctuations near the edge. It allows the fluid to slip at the boundary, accommodating the flow induced by the tangent electric field. This boundary condition introduces a compressible boundary layer that regularizes any singular edge dynamics. Additionally, the no-stress condition can be expressed as an emergent continuity equation for the edge density. The edge action corresponding to this new dynamic equation requires an additional auxiliary field at the edge, unlike the no-slip condition, which can be derived directly from variations of the hydrodynamic fields at the boundary. We show that the boundary action for this auxiliary field corresponds to a chiral boson that couples non-linearly to the condensate density evaluated at the boundary~\cite{abanov2020hydrodynamics}. Furthermore, in the presence of a tangential electric field, the chiral boson edge action must be gauged, and the no-stress condition must be modified to counterbalance the gauge anomaly forces at the boundary. 

Historically, there are two distinct approaches in the study of the FQH edge that are relevant to this work: Wen's chiral Luttinger liquid theory~\cite{wen1990compressibility, wenbook} and the study of the boundary dynamics of the CSGL action. While Wen's theory has been extensively studied, the latter has only been explored in few references, such as~\cite{nagaosa1994chern, orgad1996coulomb, orgad1997chern}. Wen's model of the FQH state starts by identifying the bulk of the FQH state with a U(1) Chern-Simons theory that lacks matter content. The edge theory is then derived by adding gapless degrees of freedom that restore the gauge invariance of the Chern-Simons theory at the boundary, naturally leading to the chiral Luttinger liquid algebra. Because the bulk lacks matter, fixing gauge invariance can be minimally achieved without introducing any edge Hamiltonian to these degrees of freedom. However, to make contact with an experimental FQH system, dynamics associated with the gapless edge are often phenomenologically added, which, in the minimal order of gradient expansion, results in the chiral Luttinger Liquid Hamiltonian. On the other hand, the CSGL theory includes the coupling between matter and the Chern-Simons fields. Therefore, fluctuations of the composite boson condensate density near the boundary fully determine the edge dynamics of the Laughlin state, without additional phenomenological parameters. 

The CSGL model with hard-wall boundary considered here differs from the ones studied previously in Refs.~\cite{nagaosa1994chern, orgad1996coulomb, orgad1997chern}, where the linearized edge dynamics is derived in the presence of a uniform and constant magnetic field and the absence of an electric field. In~\cite{nagaosa1994chern}, the authors neglected the quantum pressure, leading to an ideal fluid dynamics with Hall constraint. On the other hand, in Refs.~\cite{orgad1996coulomb, orgad1997chern}, while the quantum pressure term is retained, the authors impose the vanishing of the fluid density at the boundary. Nevertheless, in this work, we show that both approaches are incompatible with the chiral edge dynamics of the Laughlin state. 

The paper is organized as follows: we start with a brief review of the superfluid dynamics of the composite boson model in Section~\ref{sec:review}. In Section~\ref{sec:topfluid}, we introduce a variational formulation for these fluid dynamic equations, by including additional topological terms to the hydrodynamic action. In Section~\ref{sec:duality}, we demonstrate the duality between the topological fluid action and the composite boson (CSGL) action. Section~\ref{sec:bc} discusses the incompatibility of the no-slip condition with the FQH edge dynamics. In Sec.~\ref{sec:chiralboson}, we show that the anomaly equation at the edge can be derived variationally by adding a boundary chiral boson action, which is non-linearly coupled to the edge density, to the topological fluid action. In Section~\ref{sec:boundarylayer}, we offer a heuristic boundary layer interpretation of the edge anomaly equation. The paper concludes with a discussion and outlook in Section \ref{sec:concl}.

%%%%%%%%%%%%%%%%%%%%%%%%%%%%%%%%%%%%%%%%%%
\section{Composite Boson Superfluid Dynamics}
%%%%%%%%%%%%%%%%%%%%%%%%%%%%%%%%%%%%%%%%%%

\label{sec:review}

\subsection{Superfluid Equations}

In Ref.\cite{stone1990superfluid}, Michael Stone reformulates the saddle-point dynamics of composite bosons using hydrodynamic-like equations. These equations depict a dissipationless charged fluid whose vorticity is linked to density fluctuations against a constant background. Adopting a mean-field theory approach, the author assumes that the condensate density changes over scales significantly larger than the magnetic length and neglects higher derivatives of the condensate density. However, this assumption may not hold near a hard-wall boundary, where the condensate density can fluctuate markedly within distances comparable to the magnetic length, as demonstrated in~\cite{orgad1996coulomb,orgad1997chern}. Retaining the quantum pressure terms, the full fluid dynamic description of a Laughlin state with filling factor $\nu$ can be expressed in terms of the continuity equation and the Hall constraint:
\begin{align}
    \p_tn+\p_i(n \mathcal V^i)&=0\,, \la{continuity-Stone}
    \\
    \epsilon^{ij}\p_i \mathcal V_{j}+\frac{2\pi\hbar}{\nu m}n-\frac{eB}{m}&=0\,, \la{Hall-Stone}
\end{align}
as well as the Euler equation:
\begin{align}
    &\p_t \mathcal V_{i}+ \mathcal V^j\p_j\mathcal V_{i}+\frac{eB}{m}\epsilon_{ij}\mathcal V^j-\frac{1}{mn}\p_i\big(p(n)\big)=\nonumber
    \\
    &\frac{\hbar^2}{4m^2n}\,\p^j\left(\frac{\p_in\p_jn}{n}-\delta_{ij}\Delta n\right). \la{Euler-Stone}
\end{align}
Here, the velocity field is defined as $\mathcal V_i=\p_i\vartheta+a_i+\tfrac{e}{\hbar}A_i$, where $\vartheta$ is the composite boson condensate phase, $a_i$ is the statistical Chern-Simon gauge field and $A_i$ is the external electromagnetic vector potential (see Ref.~\cite{stone1990superfluid} for the full derivation). The pressure $p(n)$ is determined from the density-density interactions within the composite boson condensate. In this work, our focus is on local repulsive interactions, however, extending our analysis to non-local interactions, such as the Coulomb potential, is straightforward.

If we neglect the second line of Eq.(\ref{Euler-Stone}), the Eqs.~(\ref{continuity-Stone}-\ref{Euler-Stone}) align with the fluid dynamics equations for an inviscid, compressible, and charged fluid subject to a magnetic field. They also include an additional constitutive relation, which correlates the fluid’s vorticity with its density fluctuations. This is why the composite boson condensate is considered a superfluid. These simplified equations were the basis of Ref.~\cite{nagaosa1994chern}. However, the complete Eq.~(\ref{Euler-Stone}), being a third-order derivative equation, requires an extra boundary condition when a hard wall is present, a detail that was overlooked in their analysis.

Before discussing the potential hard-wall boundary conditions for the set of equations~(\ref{continuity-Stone}-\ref{Euler-Stone}), it is important to note that these equations are somewhat unbalanced. They represent only first-order differential equations in the velocity fields while being third-order differential equations in the density field. We have demonstrated, in Ref.~\cite{monteiro2023hamiltonian}, that this system of equations can be transformed into a system of second-order differential equations through a redefinition of the velocity field. This transformation is expressed as
\begin{equation}
\mathcal{V}^i = v^i - \frac{\hbar}{2mn} \epsilon^{ij} \partial_j n.
\end{equation}
Under this redefinition Eqs.~(\ref{continuity-Stone}) and (\ref{Hall-Stone}) become
\begin{align}
    \p_tn+\p_i(nv^i)&=0\,, \la{eq:continuity}
    \\
    \epsilon^{ij}\p_iv_j-\frac{eB}{m}+\frac{2\pi\hbar}{\nu m}n+\frac{\hbar}{2m}\p_i\left(\frac{\p^in}{n}\right)&=0\,, \la{eq:Hall}
\end{align}
whereas the Euler equation~(\ref{Euler-Stone}) writes
\be
\p_t v_i+v^j\p_j v_i=\frac{1}{m n}\p_jT^{j}_{\,\,\,i}-\frac{eB}{m}\epsilon_{ij} v^j\,,  \la{eq:Euler}
\ee
with the stress tensor $T^{j}_{\,\,\,i}$ given by
\begin{align}
T^{j}_{\,\,\,i}=&\left[p(n)+\frac{\pi\hbar^2}{\nu m}n^2\right]\delta^{j}_{\,\,i} -\frac{\hbar n}{2}(\epsilon_{ik}\p^kv^j+\epsilon^{jk}\p_iv_k)\,. \la{eq:stress}
\end{align}
The first term in Eq.~(\ref{eq:stress}) is the modified fluid pressure, whereas the second one is the odd viscosity term~\cite{avron1998odd}. The same stress tensor also shows up in the Ref.~\cite{Geracie2015-Thermal} as a result of the lowest Landau level limit $(m\rightarrow0)$ of electrons in a magnetic field.

The set of equations~(\ref{eq:continuity}-\ref{eq:stress}) is completely equivalent to Eqs.~(\ref{continuity-Stone}-\ref{Euler-Stone}). However, the former set has the advantage of being composed of second-order differential equations, both in density and velocity fields. Therefore, we will use the fluid dynamic equations (\ref{eq:continuity}-\ref{eq:stress}) as the basis for the dynamics of the full Laughlin state and as the starting point for this work. Fluid dynamic equations, in which the stress tensor includes at most first-order spatial derivatives, are commonly referred to as first-order hydrodynamics.

%%%%%%%%%%%%%%%%%%%%%%%%%%%%%%%%%%%%%%%%%%%%
\subsection{Hard wall superfluid boundary Conditions}
%%%%%%%%%%%%%%%%%%%%%%%%%%%%%%%%%%%%%%%%%%%%%

Any consistent set of boundary conditions for the superfluid hydrodynamic equations (\ref{eq:continuity}-\ref{eq:stress}) must preserve two main conservation laws: the number of electrons and the total energy of the system. Let us assume that the Hall fluid is confined within a finite, rigid domain denoted as \( \mathcal{M} \). The number of electrons inside \( \mathcal{M} \) is determined by integrating the condensate density over the entire domain. Given that the number of electrons remains constant, we conclude that
\be
\frac{d}{dt}\int_{\mathcal M}n \,d^2x=-\oint_{\p\mathcal M}n\, v_n \,ds=0\,.
\ee
This can be satisfied when the integrand vanishes, which gives us
\be
(nv_n)\Big|_{\p\mathcal M}=0\,.
\ee
Assuming the density does not vanish at the boundary, it follows that the normal component of the velocity field must be zero at the boundary. This condition, known as the no-penetration condition, implies that there is no flow of particles toward the hard wall.

The fluid energy is defined by an additional conservation law, which originates from Eqs. (\ref{eq:continuity}-\ref{eq:stress}). This equation can be expressed as:
\be
\p_t\mathcal H+\p_i(\mathcal H v^i+T^{ij}v_j)=0\,,
\ee
where the energy density is defined as 
\be
\mathcal H=\frac{m}{2}nv_i^2+V(n)\,,
\ee
and the internal energy $V(n)$ determines the modified pressure as follows
\be
p(n)+\frac{\pi\hbar^2}{\nu m}n^2=n V'(n)-V(n)\,.
\ee

The total energy of the fluid is represented by the integral of \( \mathcal{H} \) over the entire domain. For simplicity, let's assume that \( \mathcal{M} \) corresponds to the lower half-plane, meaning \( y \leq 0 \). In this scenario we have,
\be
\frac{d }{dt}\int  \mathcal H\, d^2x = -\int \left[\left(\tilde{\mathcal H}-\tilde T_{yy}\right)\tilde v_y-\tilde T_{yx}\tilde v_x\right] dx, \la{dH/dt}
\ee
where the fields evaluated at the boundary are denoted by a tilde on top, that is, $\tilde f(x, t)\equiv f(x,0,t)$. 

Energy conservation is preserved when $\tilde{v}_y = 0$ (no-penetration condition) is combined with either  $\tilde{T}_{yx} = 0$ (no-stress condition) or  $\tilde{v}_x = 0$ (no-slip condition)~\footnote{These are only sufficient conditions. The necessary conditions are beyond the scope of this work and will be addressed in future research.}.

The composite boson model of the Laughlin state can, in principle, be supplemented with different choices of boundary conditions. However, in this work, we will demonstrate that not all boundary conditions are compatible with the expected physics at the edge of the Laughlin state.

%%%%%%%%%%%%%%%%%%%%%%%%%%%%%%%%%%%%%%%%%%
\section{Topological Fluid Action}
\label{sec:topfluid}
%%%%%%%%%%%%%%%%%%%%%%%%%%%%%%%%%%%%%%%%%%%%
In this section, we will
show that the composite boson equations of motion ~Eq.~\ref{eq:continuity}, Eq.~\ref{eq:Euler}, Eq.~\ref{eq:Hall}, along with Eq.~\ref{eq:stress} can be obtained from a hydrodynamical variational principle with Clebsch potentials. It is well known that the Poisson algebra between the fluid density and the velocity field allows for the existence of Casimirs~\cite{landau1941theory, dzyalo1980poisson, holm1984relativistic, 1997-ZakharovKuznetsov, abanov2020hydrodynamics}. These Casimirs are the zero-modes of the Poisson structure since they commute with all the hydrodynamic variables. This implies that the algebra is degenerate and cannot be inverted. Consequently, it is impossible to derive a variational principle for hydrodynamics solely based on the hydrodynamic quantities. Therefore, to derive the hydrodynamic equations from an action, we must enlarge the phase space by introducing canonical variables, which remove the degeneracy of the Poisson structure. This is obtained through the introduction of three auxiliary scalar fields, named Clebsch potentials. They were first introduced by Clebsch himself in 1859 and used to parametrize the fluid velocity~\cite{clebsch1859ueber}. 

At first glance, Clebsch potentials may seem like a mere trick to derive hydrodynamic equations, but they play a systematic role in writing down topological terms in the hydrodynamic action, as it was already pointed out in some of our previous work~\cite{monteiro2015-hydro, nair2021topological}. By topological terms, we refer to action terms that do not depend on the metric and, therefore, do not contribute to the fluid stress tensor. For further reading and a comprehensive understanding of Clebsch parametrization, we refer readers to Refs.~\cite{1997-ZakharovKuznetsov,jackiw2004perfect}, as well as the references therein.

%An in-depth discussion of Clebsch parametrization is beyond the scope of this letter. For further reading and a comprehensive understanding of Clebsch parametrization, we refer readers to Refs.~\cite{1997-ZakharovKuznetsov,jackiw2004perfect}, as well as the references therein. The only information about the Clebsch potentials $(\theta, \alpha, \beta)$ that is necessary to understand this work is that they are unphysical when considered individually and should only appear in the specific combination $u_\mu \equiv \partial_\mu \theta + \alpha \partial_\mu \beta$, where $\mu = 0, 1, 2$. Detailed calculations and additional information can be found in the Supplementary Material (SM), which accompanies this letter.

%However, these Clebsch potentials are unphysical by themselves and must only appear under the combination $u_\mu=\p_\mu\theta+\alpha\p_\mu\beta$, with $\mu=0,1,2$

In general, the hydrodynamic action is a functional of the particle density $n$, the fluid velocity $v^i$, as well as the three Clebsch potentials $(\theta,\alpha,\beta)$. However, these Clebsch potentials are unphysical when considered individually and must only appear under the combination $u_\mu=\p_\mu\theta+\alpha\p_\mu\beta$, with $\mu=0,1,2$. Additionally, to obtain the Hall constraint as an additional equation of motion, we also need to include a Lagrange multiplier denoted by $b_0$. The bulk action for the ``topological fluid" considered in this work can be split into
\be
S_{\text{bulk}}=S_{\text{hydro}}+S_{\text{top}}\,, \la{action-bulk}
\ee
where $S_{\text{hydro}}$ refers to the ordinary hydrodynamic bulk action when we set $b_0=0$, and $S_{\text{top}}$ contains the aforementioned topological terms. The explicit form of $S_{\text{hydro}}$ is given by
\begin{align}
&S_{\text{hydro}}=-\int\left[\hbar\int n \left(u_0+b_0\right)d^2x+H\right] dt,  \la{S-hydro}
\\
&H=\int d^2x\left[\hbar nv^iu_i-\frac{m}{2}nv_i^2+V(n)+\frac{\hbar}{2}\epsilon^{ij}v_i\p_j n\right], \la{H-hydro}
\end{align}
%\begin{align}
%S_{\text{hydro}}=&\,-\int d^3x\left[\hbar n (u_0+a_0+v^iu_i)-\tfrac{1}{2}mnv_i^2\right.\nonumber
%\\
%&-\left.V(n)-\iota_H\epsilon^{ij}v_i\p_j n\right], \la{S-hydro}
%\end{align}
where $m$ is the electron effective mass, $V(n)$ is the fluid internal energy (as defined in the previous section) and $\hbar$  was introduced for dimensional reasons.

The topological terms in $S_{\text{top}}$ can be written as
\begin{align}
S_{\text{top}}&=\frac{\nu}{2\pi}\int \left[e u\wedge dA+ b_0 \,dt\wedge\left(e\, dA-\hbar \,du\right)\right], \la{S-top}
\\
&=\frac{\nu}{2\pi}\int d^3x \left[eB\left(u_0+b_0\right)+\epsilon^{ij}\left(eu_iE_j-\hbar b_0\p_iu_j\right)\right], \nonumber
\end{align}
where $A_\mu$ is the electromagnetic potential, $E_j$ the electric field, $B$ the magnetic field, $e$ is the elementary charge and $\nu$ is the filling factor. Here, $\nu$ is taken to be the inverse of an odd whole number. When $E_j=0$, the action $S_{\text{bulk}}$ is identical to the one considered in Ref.~\cite{nair2021topological}, with the specific choice of the Hamiltonian as in Eq.~(\ref{H-hydro}).

\subsection{Polarization field} 
\label{sec:pol}
Before turning our attention to the fluid dynamic equations derived from the action~(\ref{action-bulk}-\ref{S-top}), note that the term proportional to the electric field $E_j$ in $S_{\text{top}}$ enables us to identify the fluid polarization as 
\begin{align}
	P^i&=-\frac{e\nu}{2\pi}\epsilon^{ij}u_j.  \la{Polarization}
\end{align}

This implies that both the superfluid velocity and the condensate density can be parametrized by the polarization field. The equation of motion for the velocity field $v^i$ provides its parametrization in terms of Clebsch potentials, that is,
\be
v_i=\frac{\hbar}{m}\left(u_i+\frac{\epsilon_{ij}}{2n}\p^jn\right)=\frac{\hbar}{m}\epsilon^{ij}\left(\frac{2\pi P_i}{\nu e}+\frac{\p_jn}{2n}\right), \la{velocity}
\ee
where we used Eq.~(\ref{Polarization}) in the second equality. Additionally, varying the action $S_{\text{bulk}}$ over the Lagrange multiplier $b_0$ imposes the constraint
\be
n=\frac{\nu eB}{2\pi\hbar}-\frac{\nu}{2\pi}\epsilon^{ij}\p_iu_j=\frac{\nu eB}{2\pi\hbar}+\frac{1}{e}\p_iP^i\,. \la{Hall-constraint}
\ee
In classical electromagnetism, the total charge in a conductor is divided into free charge and bound charge densities, where the latter is often expressed in terms of the divergence of the polarization density. Therefore, Eq.~(\ref{Hall-constraint}) allows us to identify $\nu eB/(2\pi\hbar)$ with the free charge density in the sample. 

The polarization field provides a complete set of observables. This suggests an alternate description with an action directly in terms of $P^i$, whose Poisson brackets are proportional to the Hall conductivity, that is,
\begin{align}
\{ P^i({\boldsymbol x}), P^j ({\boldsymbol x'})\} &= -\frac{\nu e^2}{2 \pi \hbar} \epsilon^{ij} \delta( {\boldsymbol x} - {\boldsymbol x'}). \la{polarization-bracket}
\end{align}
Calculation details of the polarization algebra starting from the Poisson structure can be found in Appendix A.
Using that $u_i=\frac{2\pi}{\nu e}\epsilon_{ij}P^j$, we find that the bulk action $S_{\text{bulk}}$ can be expressed as
\be
S_{\text{bulk}}= \int  \left({\pi \hbar \over \nu e^2} \epsilon_{ij} P^i \partial_t P^j
-  H\right)d^3x\,,
\ee
where the first term in this action leads to the polarization algebra and all the factors of $u_i$, $v^i$, and $n$ in $H$ given in Eq.~\ref{H-hydro} must be expressed through the following replacements
\begin{align}
 &u_i\rightarrow\frac{2\pi}{\nu e}\epsilon_{ij}P^j \nonumber\\
 &v^i\rightarrow \frac{\hbar}{m}\epsilon^{ij}\left(\frac{2\pi P_j}{\nu e}+\frac{\p_jn}{2n}\right)\nonumber\\ 
 &n\rightarrow\frac{\nu e B}{2\pi\hbar}+\frac{1}{e}\p_iP^i \nonumber
\end{align}
The above Hamiltonian along with the Polarization algebra can serve as a starting point of canonical quantization and will be considered in a separate publication. 

\subsection{Bulk hydrodynamic equations} 

In the previous section, it was mentioned that the variation of the action with respect to $v^i$ gives us the Clebsch parametrization of the velocity field~(\ref{velocity}), while the action variation with respect to $b_0$ yields the constraint~(\ref{Hall-constraint}). By combining both, we obtain our first hydrodynamic equation — the Hall constraint~(\ref{eq:Hall}):
\[
\epsilon^{ij}\p_iv_j-\frac{eB}{m}+\frac{2\pi\hbar}{\nu m}n+\frac{\hbar}{2m}\p_i\left(\frac{\p^in}{n}\right)=0\,.
\]
On the other hand, the continuity equation~(\ref{eq:continuity}) arises directly from the bulk variation of the Clebsch potential $\theta$, after imposing Faraday's law.

Obtaining the Euler equation starting from the topological fluid action is more involved, as it does not directly follow from one or two equations of motion alone, but rather from a combination of all of them. Therefore, let us focus on the remaining equations of motion, which are derived from variations of the fields $(n, \alpha, \beta)$. They can be expressed as follows:
\begin{align}
    &\delta n: \quad u_0+b_0+v^iu_i-\frac{m}{2\hbar}v_i ^2+\frac{V'(n)}{\hbar}+\frac{\epsilon^{ij}}{2}\p_iv_j=0\,,
    \\
    &\delta\alpha: \,\, \left[\epsilon^{ij}\left(\p_tu_j-\p_j(u_0+b_0)+\frac{e}{\hbar}E_j\right)-\frac{2\pi}{\nu}nv^i\right]\p_i\beta=0, \la{eom-alpha}
    \\
    &\delta\beta: \,\, \left[\epsilon^{ij}\left(\p_tu_j-\p_j(u_0+b_0)+\frac{e}{\hbar}E_j\right)-\frac{2\pi}{\nu}nv^i\right]\p_i\alpha=0. \la{eom-beta}
\end{align}
For calculation details, please refer to appendix~\ref{app:eom}.

Since $\alpha$ and $\beta$ are assumed to be independent variables, combining Eqs.~(\ref{eom-alpha}) and (\ref{eom-beta}) implies that
\be
\p_tu_i-\p_i(u_0+b_0)+\frac{e}{\hbar}E_i-\frac{2\pi n}{\nu}\epsilon_{ij}v^j=0\,.
\ee
After some algebraic manipulations, this equation can be brought to the form
\be
\p_t v_i+v^j\p_j v_i=\frac{1}{m n}\p_jT^{j}_{\,\,\,i}-\frac{e}{m}(E_i+B\epsilon_{ij} v^j),  \la{Euler}
\ee
where the stress tensor on the right hand side is defined in Eq.~(\ref{eq:stress}). Note that, this equation only differs from Eq.~(\ref{eq:Euler}) by the presence of electric field.

\section{Duality between CSGL and Topological fluid action} 
\label{sec:duality}

So far, the connection between the CSGL theory and fluid dynamics has been established at the level of equations of motion. However, these two models are also connected through a duality transformation. For simplicity, let us consider the fluid to be spread over the whole plane to avoid the complexities arising from boundaries in the domain. For a complete description of this duality, which carefully takes into account the presence of boundaries, we direct the reader to the appendix~\ref{app:duality}.  

In this context, we can neglect boundary terms and the topological action $S_{\text{top}}$ can be written as 
\begin{align}
S_{\text{top}}=&-\frac{\hbar\nu}{4\pi}\int \left(b_0dt+\alpha d\beta-\tfrac{e}{\hbar} A\right)\wedge d\left(b_0dt+\alpha d\beta-\tfrac{e}{\hbar} A\right)\nonumber
\\
&\,+\frac{\nu e^2}{4\pi\hbar}\int A\wedge dA\,. \la{S-top-dual}
\end{align} 
Note that the first line of Eq.~(\ref{S-top-dual}) becomes the Chern-Simons action, as long as we are able to parametrize the gauge field $a_\mu$ as 
\be
a= b_0dt+\alpha d\beta-\frac{e}{\hbar} A+d\lambda\,. \la{a-parametrization}
\ee
Here, the term $d\lambda$ was included for generality, since it only contributes to boundary terms. 

It is worth pointing out that, because of the term $b_0 dt$, the gauge field parametrization used here is different from the ones employed in Refs.~\cite{deser2001clebsch, tong2022gauge}, where the incompleteness of the Clebsch decomposition of Chern-Simons and Maxwell's actions were highlighted.

The gauge field parametrization~(\ref{a-parametrization}) can be imposed through a Lagrange multiplier, which allows us to express $S_{\text{bulk}}$ in the form
\begin{align}
S_{\text{bulk}}[\zeta]&=S_{\text{hydro}}[a_\mu]-\frac{\hbar\nu}{4\pi}\int a\wedge da +\frac{\nu e^2}{4\pi\hbar}\int A\wedge dA \nonumber
\\
&+\int \zeta\wedge d\left(a+\frac{e}{\hbar}A-b_0dt+\alpha d\beta\right), \la{S-zeta}
\end{align}
where, the action $S_{\text{hydro}}[a_\mu]$ is obtained from $S_{\text{hydro}}$ through the replacement the terms $u_0+b_0$ and $u_i$ by $\p_t\vartheta+a_0+\frac{e}{\hbar}A_0$ and $\p_i\vartheta+a_i+\frac{e}{\hbar}A_i$, respectively. Upon integrating out the Lagrangian multiplier $\zeta_\mu$ and denoting the combination $\vartheta+\lambda$ by $\theta$, the action $S_{\text{bulk}}[\zeta]$ reduces to the topological fluid action $S_{\text{bulk}}$ from Eqs.~(\ref{action-bulk}-\ref{S-top}). 

The variables $(b_0,\alpha,\beta)$ only appear linearly in the action $S_{\text{bulk}}[\zeta]$ and integrating them out impose some restrictions on the Lagrange multiplier $\zeta_\mu$. The action variations with respect to them gives us the following:
\begin{align}
    &\delta b_0:\quad  \epsilon^{ij}\p_i\zeta_j=0\,, \la{eom-b0-zeta}
    \\
    &\delta \alpha:\quad  \epsilon^{ij}\left[\p_i\zeta_j\p_t\beta+(\p_t\zeta_i-\p_i\zeta_0)\p_j\beta\right]=0\,, \la{eom-alpha-zeta}
    \\
    &\delta \beta:\quad  \epsilon^{ij}\left[\p_i\zeta_j\p_t\alpha+(\p_t\zeta_i-\p_i\zeta_0)\p_j\alpha\right]=0\,. \la{eom-beta-zeta}
\end{align}
These equations of motion impose that $\zeta_\mu=\p_\mu\chi$, and plugging this expression for $\zeta_\mu$ back into the action $S_{\text{bulk}}[\zeta]$, we find that the second line of Eq.~(\ref{S-zeta}) becomes a total derivative. Therefore, upon this substitution, we are left with an action solely in terms of the fields $(n,\vartheta, v^i, a_\mu)$. To see that this action is indeed the CSGL action, let us now turn our attention to $S_{\text{hydro}}[a_\mu]$ and combine the Madelung variables $(n,\vartheta)$ into the bosonic scalar field $\Phi=\sqrt{n}\,e^{i\vartheta}$. After some algebra, we obtain
\begin{align}
S_{\text{hydro}}[a_\mu]&=\int\left[\Phi^\dagger\left(i\hbar D_t+\frac{i\hbar}{2} v^i D_i +\frac{m}{2}v_i^2\right)\Phi-V(|\Phi|)\right. \nonumber
\\
&-\left.\frac{i\hbar}{2}v^i (D_i\Phi)^\dagger\,\Phi+\frac{\hbar}{2}\epsilon^{ij}v_i\p_j(|\Phi|^2)\right]d^3x\,,
\end{align}
where, $D_\mu\equiv\p_\mu+i\left(a_\mu+\tfrac{e}{\hbar}A_\mu\right)$ denotes the covariant derivative. Integrating out the velocity field in $S_{\text{hydro}}[a_\mu]$ gives us 
\be
S_{\text{bulk}}[\zeta]\rightarrow S_{\text{CSGL}}+\frac{\nu e^2}{4\pi\hbar}\int A\wedge dA, \la{bulk-CSGL}
\ee
with
\begin{align}
S_{\text{CSGL}}&=\int d^3x\left[i\hbar\Phi^\dagger D_t\Phi-\frac{\hbar^2}{2m}|D_i\Phi|^2-V(|\Phi|)\right.\nonumber
\\
&\left.-\frac{|\Phi|^2}{m}\left(\hbar\epsilon^{ij}\p_ia_j+eB\right)-\frac{\hbar\nu}{4\pi}\epsilon^{\mu\lambda\kappa} a_\mu\p_\lambda a_\kappa\right]. \la{S-CSGL}
\end{align}

The action~(\ref{S-CSGL}) is the same as the one obtained in the Ref.~\cite{abanov2013effective}, if we set its phenomenological parameter $\xi$ to be $\frac{1}{2}\hbar$. There, $-\xi$ refers to the Hall viscosity of the FQH state, whereas, here $-\xi$ represents the odd viscosity coefficient of the topological fluid. It is important to note that, in principle, the Hall viscosity of the FQH state and the odd viscosity of the topological fluid need not have the same value. A proper definition of the Hall viscosity will require coupling the action~(\ref{S-CSGL}) to a strain rate or a time-dependent metric~\cite{avron1995viscosity, read2009non,haldane2011geometrical,bradlyn2012kubo, hoyos2012hall, hughes2013torsional, hoyos2014hall, laskin2015collective, can2014fractional,can2015geometry,klevtsov2015geometric,klevtsov2015quantum, gromov2014density, gromov2015framing, gromov2016boundary,karabali2016geometry,karabali2016role} before mapping it to a hydrodynamic system.

This shows that both the topological fluid action $S_{\text{bulk}}$ and Chern-Simon-Ginzburg-Landau action $S_{\text{CSGL}}$ can be derived from a more general action principle $S_{\text{bulk}}[\zeta]$, demonstrating the duality between them. This duality holds true even in the presence of boundaries, as we show in the appendix~\ref{app:duality}. However, in that case, the boundary terms ignored in this section make the analysis slightly more complicated.

\section{Incompatibility of no-slip boundary conditions with edge dynamics}
\label{sec:bc}

To incorporate boundary effects, we consider the simpler scenario where the Hall fluid is confined to the lower half-plane, that is, $y\leq0$. As discussed in Sec.~\ref{sec:review}, in the absence of electric fields, one would typically expect the boundary conditions at a hard wall to consist of the no-penetration condition $(\tilde{v}_y=0)$ along with either the no-slip $(\tilde{v}_x=0)$ or no-tangential stress $(\tilde{T}_{yx}=0)$ boundary conditions. Again, we are denoting fields evaluates at the boundary with a tilde on top. However, the presence of the gauge anomaly at the edge of the domain is expected to modify the no-penetration condition, indicating an influx of particles from the bulk to the edge due to a nonvanishing tangent electric field at the boundary. This is referred as the anomaly inflow mechanism, where the tangent electric field drives an electric current normal to the boundary.

From a variational principle viewpoint, boundary conditions can be obtained from action variations with respect to the dynamical fields evaluated at the boundary. Therefore, the anomaly inflow condition is precisely what we obtain from the topological fluid action $S_{\text{bulk}}$ if we allow the variation of the field $\theta$ to be arbitrary at the boundary. In other words, the no-penetration condition must be replaced by
\be
\tilde n \tilde v_y+\frac{\nu e}{2\pi\hbar}\tilde E_x=0\,. \la{anomaly-inflow}
\ee
in the presence of a tangent electric field. As expected, the boundary condition~(\ref{anomaly-inflow}) reduces to the no-penetration condition $(\tilde{v}_y=0)$ if we set $\tilde E_x=0$.

The velocity field $v^i$ and the Lagrangian multiplier $b_0$ do not generate any other boundary condition. Nevertheless, variations of the fields $n$, $\alpha$, and $\beta$ at the boundary, we obtain three additional expressions:
\begin{align}
    &\delta \tilde n:\quad \tilde v_x=0\,, \la{no-slip}
    \\
    &\delta \tilde \alpha:\quad \tilde b_0\, \p_x\tilde \beta=0\,,  \la{b0-beta}
    \\
    &\delta \tilde \beta:\quad \tilde b_0\, \p_x\tilde \alpha=0\,. \la{b0-alpha}
\end{align}
Note that the no-slip condition~(\ref{no-slip}) arises directly from the action $S_{\text{bulk}}$ and that $\tilde b_0=0$ is a particular solution of Eqs.~(\ref{b0-beta}) and (\ref{b0-alpha})~\footnote{In this case, going back to the action $S_{\text{bulk}}$, we note that there is no canonical momentum $\Pi_0$ for $b_0$. Therefore, we can choose $b_0 \approx 0$ as the constraint conjugate to $\Pi_0$ in carrying out a canonical reduction. This means that $b_0 = 0$ can be obtained as a gauge choice even in the bulk, leading to $\tilde b_0=0$ even without the variational procedure.}.

To check the interplay between the anomaly inflow and the no-slip condition, we integrate the continuity equation from $-\ell$ to $+\ell$ and take the limit $\ell\rightarrow 0$:
\be
\p_t\left(\lim_{\ell\rightarrow0^+}\int\limits_{-\ell}^0n\,dy\right)+\p_x\left(\lim_{\ell\rightarrow0^+}\int\limits_{-\ell}^0nv_x\,dy\right)=-\frac{\nu e}{2\pi\hbar}\tilde E_x\,. \la{incompatibility}
\ee
Here we used that $n=0$ for $y>0$, by assumption. For $\tilde v_x=0$, the LHS of Eq.~(\ref{incompatibility}) leads to a singular density profile with no flow of charge along the edge. This is in contrast with the FQH physics, where we expect anomaly inflow-induced chiral edge dynamics. 

In the following, we proceed to investigate whether the no-tangent stress boundary condition $\tilde T_{yx}=0$ can provide the necessary chiral edge dynamics consistent with the anomaly inflow mechanism~(\ref{anomaly-inflow}).

\section{No-stress boundary condition and chiral boson action} 
\label{sec:chiralboson}

In the absence of an electric field, we can derive $\tilde T_{yx}=0$ from a variational principle by introducing an auxiliary boundary field $(\phi)$. This is only the case because $\tilde T_{yx}=0$ is a dynamical equation in disguise. By using both the continuity equation and the no-penetration condition $\tilde v_y=0$, we can show that
\begin{align}
	\tilde T_{yx}&=\frac{\hbar n}{2} (\partial_x v_x-\partial_y v_y)\Big|_{y=0}\nonumber\\&=\hbar \sqrt{\tilde n}\left[\partial_t(\sqrt{\tilde n})+\p_x(\sqrt{\tilde n}\,\tilde v_x)\right],\la{no-stress-dyn}	
\end{align}
In fact, the dynamical equation~(\ref{no-stress-dyn}) represents the equation of motion for the auxiliary field $\phi$. The appropriate boundary action can be obtained by taking the hard wall limit of the free-surface action discussed in Ref.~\cite{abanov2020hydrodynamics}. The resulting edge action takes the form
\begin{align}
	S_{\text{edge}}=\frac{\hbar}{2}\int dt \, dx\, \p_t\phi(\p_x\phi-2\sqrt{\tilde n}).\la{ungaugededge}
\end{align}

As pointed out in Ref.~\cite{abanov2020hydrodynamics}, this boundary action takes the form of a chiral boson coupled to the ``edge density" $\sqrt{\tilde n}$. Following this comparison, we can couple the field $\phi$ to the electromagnetic gauge field in the same way as the chiral boson field, which leads to 
 \be
S_{\text{edge}}=\frac{\hbar}{2}\int dt  dx\,\bigg(\p_t\phi+\frac{\nu e \tilde A_0}{2\pi\hbar}\bigg)\bigg(\p_x\phi+\frac{ \nu e \tilde A_x}{2\pi\hbar}-2\sqrt{\tilde n}\bigg). \la{action-edge}
\ee

The addition of $S_{\text{edge}}$ preserves Eqs.~(\ref{anomaly-inflow}-\ref{b0-alpha}), but replaces the no-slip condition~(\ref{no-slip}) with the following dynamical equations for the boundary fields:
 \begin{align}
\p_t\phi+\frac{ \nu e}{2\pi\hbar}\tilde A_0&=-\sqrt{\tilde n}\,\tilde v_x\,,  \la{bosonization}
\\
\p_t(\sqrt{\tilde n})+\p_x(\sqrt{\tilde n}\,\tilde v_x)&=-\frac{\nu e}{4\pi\hbar}\tilde E_x\,. \la{edge-continuity}
\end{align}
Equation~(\ref{edge-continuity}) can be interpreted as the gauge anomaly equation when considering the following identification:
\be
\rho =-e\sqrt{\tilde n}\qquad \text{and} \qquad  I =-e\sqrt{\tilde n}\,\tilde v_x\,, \la{identification}
\ee
where $\rho$ and $I$ represent the edge charge density and the edge current, respectively. Moreover, Eq.~(\ref{bosonization}) together with the identification~(\ref{identification}) is resembles the bosonization expression for the edge current, that is,
\be
I =e\left(\p_t\phi+\frac{\nu e}{2\pi \hbar}\tilde A_0\right). 
\ee

Furthermore, by combining Eqs.~(\ref{eq:continuity}) and (\ref{edge-continuity}), we observe that the gauge anomaly induces tangential stresses on the wall in the presence of an electric field, which are given by
\begin{align}
\tilde T_{yx}&=-\frac{\nu e}{4\pi}E_x\Big(\sqrt{n}+\frac{\p_y n}{n}\Big)\Big|_{y=0}. \la{no-stress}
\end{align} 
Thus, the complete hydrodynamic action for the topological fluids, including the bulk and boundary contributions for the domain $y \leq 0$, consistent with the anomaly inflow condition follows from
\begin{align}
S = S_{\text{hydro}} + S_{\text{top}} + S_{\text{edge}}\,. \la{action}
\end{align}

\section{Boundary layer}
\label{sec:boundarylayer}
Here, we present a heuristic argument that suggests how the consistent chiral edge dynamics can be achieved by considering a particular boundary layer regularization of Eq.~(\ref{incompatibility}). To regulate this equation, we adopt a fluid dynamic viewpoint where the boundary layer is treated as infinitesimally thin in the long wavelength regime of the fluid. This means that the thickness of the boundary layer is much smaller than any characteristic wavelength of the problem.

In an FQH sample, fluctuations usually occur at length scales much larger than the magnetic length $\ell_B\equiv\sqrt{\hbar/e B}$. Therefore, let us consider the boundary layer thickness $\ell$ to be finite and of the same order of magnitude as the magnetic length $\ell_B$. Assuming that all fluid dynamic variables vary slowly outside the boundary layer, we can examine the Hall constraint~(\ref{eq:Hall}) at $y=-\ell$, which yields the following expression:
\be
\frac{1}{\ell_B^2}=\frac{2\pi }{\nu}\tilde n^{(\ell)}+\p_i\left(\frac{\p^i\tilde n^{(\ell)}}{2\tilde n^{(\ell)}}\right)+\frac{ m}{\hbar}\epsilon^{ij}\p_i\tilde v^{(\ell)}_j\approx \frac{2\pi }{\nu}\tilde n^{(\ell)}\,,
\ee
where, for brevity, we denoted $n(x,-\ell,t)$ and $v_x(x,-\ell,t)$ by $\tilde n^{(\ell)}$ and $\tilde v_x^{(\ell)}$, respectively. Choosing $\ell=\ell_B\sqrt{8\pi/\nu}$ and integrating a general hydrodynamic quantity $f(x,y,t)$ over the boundary layer, we find that
\be
\int\limits_{-\ell}^0 f\,dy= \tilde f^{(\ell)}\,\ell+O(\ell^2)=\frac{2 \tilde f^{(\ell)}}{\sqrt{\tilde n^{(\ell)}}} +O(\ell_B^2)\,. 
\ee
This approximation provides a way to regularize Eq.~(\ref{incompatibility}). Hence, at leading order in the magnetic length, we end up with
\be
\p_t\sqrt{\tilde n^{(\ell)}}+\p_x\left(\sqrt{\tilde n^{(\ell)}} \,\tilde v_x^{(\ell)}\right)=-\frac{\nu e \tilde E_x}{4\pi\hbar}\,. \la{gauge-anomaly-reg}
\ee

As the magnetic field becomes more intense $(\ell_B\rightarrow 0)$, the boundary layer becomes smaller and smaller. Consequently, Eq.~(\ref{gauge-anomaly-reg}) tends to Eq.~(\ref{edge-continuity}). In a sense, we can consider Eq.~(\ref{edge-continuity}) as the effective dynamical boundary condition for the FQH state, obtained by integrating out the boundary layer~\footnote{For details of a closely related compressible boundary layer profile that arises due to $\tilde T_{xy}=0$ for neutral fluids, we refer to Ref.~\cite{abanov2020hydrodynamics}.}.

%Finally, we have generalized this duality to include the boundary and the details of this calculation are given in the supplementary information.~\cite{SM}. Finally, the full flux-attachment action coming from the FQH topological fluid is given by $S_{\text{bulk}}+S_{\text{edge}}$, where $S_{\text{edge}}$ is the action in Eq.~(\ref{action-edge}) with the replacement $\sqrt{\tilde n}\rightarrow|\tilde\Phi|$. This density coupling is somewhat unusual from the chiral Luttinger liquid point of view, but it arises naturally in the hydrodynamic framework. Also, differently from the TQFT phenomenology, in the CSGL theory derived from the topological fluid action, the statistical field $a_\mu$ do not couple directly to the chiral boson at the edge. 

%In the presence of a boundary, Eq.~(\ref{bulk-CSGL}) gets modified by some edge terms, that is, 
%\begin{align}
%S_{\text{bulk}}&=S_{CSGL}-\frac{\nu e^2}{4\pi\hbar}\int A\wedge dA-\frac{i\hbar}{m}\int \tilde\Phi^\dagger \tilde D_x\tilde\Phi\, dt\,dx\nonumber
%\\
%&+\frac{\nu}{4\pi}\int\left[\text{Im}\left( d\ln\tilde \Phi\right)\wedge(\hbar\tilde a-e\tilde A)+e\tilde a\wedge\tilde A\right].
%\end{align} 
%Here, the electromagnetic Chern-Simon term is necessary to ensure gauge invariance of $S_{\text{bulk}}$ as well as responsible for the anomaly inflow mechanism.

\section{Discussion and Outlook} 
\label{sec:concl}
In conclusion, this work proposes using a topological fluid dynamics action as an alternative method for studying Laughlin states. Additionally, we demonstrate that the topological fluid action can be mapped onto a Chern-Simons-Ginzburg-Landau theory through a duality transformation.

As with any theory, the requirement for a well-defined variational formulation dictates the permissible set of boundary conditions. We show that the anomaly inflow mechanism replaces the no-penetration boundary condition in the presence of an external electric field tangential to the boundary. The no-slip boundary condition is excluded due to the absence of chiral dynamics along the edge caused by the anomaly. In contrast, the no-stress boundary condition leads to chiral dynamics at the boundary, regulated by the compressible boundary layer mechanism. This no-stress boundary condition can be derived from an edge action involving an additional auxiliary chiral boson field coupled with the matter density at the boundary.

In Wen's theory~\cite{wen1990compressibility}, the description of the FQH bulk relies solely on the Chern-Simons (CS) action, which is a topological field theory with a vanishing Hamiltonian. However, when boundaries are introduced, the CS theory loses its gauge invariance. To restore this symmetry, a chiral boson field is added to the boundary. While the CS term in the bulk determines the chiral boson algebra, the chiral boson Hamiltonian is often introduced {\it a posteriori} to generate edge dynamics~\cite{wen1990compressibility}. 

In contrast, our work employs a gauge invariant hydrodynamic framework of the composite boson CSGL theory. It is important to note that this fluid perspective of the Laughlin state goes beyond a mere reinterpretation of previous results. Instead, it allows us to systematically derive non-linear edge dynamics of the composite boson model of the Laughlin state ~\cite{monteiro2021nonlinear}. Therefore, our work paves the way for studying the fluid aspects of the FQH state beyond topological quantum field theories and, equivalently, the edge dynamics beyond the chiral Luttinger liquid theory.

%In fact, other known examples of quantum fluids can be characterized by hydrodynamic actions endowed with topological terms. A general question of interest is whether the classification of topological terms can yield a complete set of possible quantum fluids.

%Our CSGL model with hard-wall boundary differs from the ones studied previously in Refs.~\cite{nagaosa1994chern, orgad1996coulomb, orgad1997chern}, where the linearized edge dynamics is derived in the presence of a uniform and constant magnetic field and in the absence of an electric field. In~\cite{nagaosa1994chern}, the authors neglected the quantum pressure, leading to an ideal fluid dynamics with Hall constraint. Within this approximation, the only possible boundary condition is the no-penetration condition. On the other hand, in Refs.~\cite{orgad1996coulomb, orgad1997chern}, while the quantum pressure term is retained, the authors impose the
%vanishing of the fluid density at the boundary. In fact, the boundary condition $\tilde n=0$, used in the Refs.~\cite{orgad1996coulomb, orgad1997chern}, can be obtained by adding the boundary action $-\frac{\hbar}{2}\int dtdx\, \tilde n \tilde v_x$ to $S_{\text{hydro}}$ in Eq.~(\ref{S-hydro}). Nevertheless, the condensate density vanishing at the wall is obviously incompatible to the anomaly inflow~(\ref{anomaly-inflow}), making the variational principle inconsistent. Lastly, the hydrodynamic theory derived here can be used to determine the nonlinear edge dynamics, as it will be addressed in a future work. 

%\section{Acknowledgments}
%%%%%%%%%%%%%%%%%%%%%%%%%%%%%%%%%%
\begin{acknowledgments}
We would like to thank Alexander Abanov for fruitful discussions and earlier collaborations. GMM was supported by the National Science Foundation under Grant OMA1936351. VPN was supported in part by NSF Grants Nos. PHY-2112729 and PHY-1820271. SG is supported by NSF CAREER Grant No. DMR-1944967 (SG).
\end{acknowledgments}

%%%%%%%%%%%%%%%%%%%%%%%%%%%%%%%%
%%%%%%%%%%%%%%%%%%%%%%%%%%%%%%%%
%\bibliographystyle{my-refs}
%\bibliography{oddviscosity-bibliography.bib}

\appendix
\onecolumngrid

\section{Polarization algebra and effective action}

For simplicity, let us consider the fluid domain to be the whole plane and ignore boundary subtleties. The term in the bulk action $S_{\text{bulk}}$ which determines the
canonical structure is
\be
\Theta=\int\left(\frac{\nu e B}{2\pi}-\hbar n\right)u_0\,d^2x=\int\left(\frac{\nu e B}{2\pi}-\hbar n\right) ({\dot \theta} + \alpha {\dot \beta})\,d^2x\equiv \int\left(\Pi_\theta\dot\theta+\Pi_\beta\dot\beta\right)d^2x\,. \la{S1}
\ee
From the canonical Poisson brackets
\begin{align}
&\{\theta(\boldsymbol x),\Pi_\theta(\boldsymbol x')\}=\{\beta(\boldsymbol x),\Pi_\beta(\boldsymbol x')\}=\delta(\boldsymbol x-\boldsymbol x')\,, \la{S3}
\\
&\{\theta(\boldsymbol x),\beta(\boldsymbol x')\}=\{\theta(\boldsymbol x),\Pi_\beta(\boldsymbol x')\}=\{\beta(\boldsymbol x),\Pi_\theta(\boldsymbol x')\}=0\,, \la{S4}
\end{align}
we are able to determine the following Poisson algebra between Clebsch potentials:
\be
\{ \theta ({\boldsymbol x}), \alpha ({\boldsymbol x'})\} = -\frac{\alpha}{\Pi_\theta} \,\delta({\boldsymbol x} -
{\boldsymbol x'})\,,\qquad \{ \alpha ({\boldsymbol x}), \beta ({\boldsymbol x'}) \} = -\frac{1}{\Pi_\theta} \delta ({\boldsymbol x} - {\boldsymbol x'})\,. \la{S5}
\ee

The calculation of the Poisson bracket of the polarization~(\ref{polarization-bracket}) is now straightforward and leads to
\begin{align*}
\{ P^i({\boldsymbol x}), P^j ({\boldsymbol x'})\} &= - {\nu e^2 \over 2\pi\hbar} \epsilon^{ij}
\frac{\hbar\nu}{2\pi\Pi_\theta}\epsilon^{kl} \partial_k \alpha
\partial_l \beta\,\delta( {\boldsymbol x} - {\boldsymbol x'}) = -{\nu e^2  \over 2 \pi \hbar} \epsilon^{ij} \delta( {\boldsymbol x} - {\boldsymbol x'})\,,
\end{align*}
where the Hall constraint~(\ref{Hall-constraint}) was imposed in the second equality. Note that the Hall constraint is compatible with the fundamental brackets, since
\be
\{P^i({\boldsymbol x}),\Pi_\theta(\boldsymbol x')\}= \frac{\nu e}{2\pi}\epsilon^{ij}\{\p_j\theta({\boldsymbol x}),\Pi_\theta(\boldsymbol x')\}={\nu e  \over 2 \pi} \epsilon^{ij} \frac{\p}{\p x^j}\delta( {\boldsymbol x} - {\boldsymbol x'})=-\frac{\hbar}{e}\{ P^i({\boldsymbol x}), \p_jP^j ({\boldsymbol x'})\}. 
\ee
This shows that it is possible to
impose the Hall constraint in the strong sense for the set of variables
$(n, P^i )$. Because the density $n$ and the velocity field $v^i$ can be written in terms of $P^i$, the configuration space of the fluid is completely determined by the polarization field, that is, the relevant Poisson brackets involving
$n$ and $v^i$ can be worked out from the polarization algebra by defining $n$ and $v^i$ in terms of 
$\partial_i P^i$. 

Given that the Hall constraint can be imposed strongly,
we may seek an action in terms of $P^i$ which directly leads to 
its Poisson bracket. This alternative description can be obtained by rewriting $\Pi_\theta=\frac{\hbar\nu}{2\pi}\epsilon^{ij}\p_i u_j$ into Eq.~(\ref{S1}) and noting that
\be
\Theta=\frac{\hbar\nu}{2\pi}\int\epsilon^{ij}u_0\p_iu_j \,d^2x=\frac{\hbar\nu}{4\pi}\int\epsilon^{ij}u_i\p_tu_j \,d^2x +\text{total derivatives}\,.
\ee

Using that $u_i=\frac{2\pi}{\nu e}\epsilon_{ij}P^j$, we find that the bulk action $S_{\text{bulk}}$ can be expressed as
\be
S_{\text{bulk}}= \int  \left({\pi \hbar \over \nu e^2} \epsilon_{ij} P^i \partial_t P^j
- \mathcal H\right)d^3x\,,
\ee
where the first term in this action leads to the polarization algebra and all the factors of $v^i$ and $n$ in $\mathcal H$ must be expressed through the following replacements
\be
v^i\rightarrow \frac{\hbar}{m}\epsilon^{ij}\left(\frac{2\pi P_j}{\nu e}+\frac{\p_jn}{2n}\right),\qquad\text{and}\qquad n\rightarrow\frac{\nu e B}{2\pi\hbar}+\frac{1}{e}\p_iP^i\,.
\ee

%%%%%%%%%%%%%%%%%%%%%%%%%%
 \section{Hydrodynamic Equations of Motion}
 \la{app:eom}
 
In this section, we will derive the hydrodynamic equations as well as the appropriate boundary conditions for the FQH fluid described by our topological fluid action. Following the notation in the main text, we have that the full action is given by $S=S_{\text{bulk}}+S_{\text{edge}}$, where the boundary term $S_{\text{edge}}$ does not contribute to bulk hydrodynamic equations. Thus, let us ignore $S_{\text{edge}}$ for now and focus on the bulk action, that is,
\[
S_{\text{bulk}}=-\hbar\int \left[\left(n-\frac{\nu e B}{2\pi\hbar}\right)(u_0+b_0)+\left(nv^i-\frac{\nu e }{2\pi\hbar}\epsilon^{ij}E_j\right)u_i-\frac{m}{2\hbar}nv_i^2+\frac{V(n)}{\hbar}+\frac{\epsilon^{ij}}{2}v_i\p_j n+\frac{\nu}{2\pi}b_0\epsilon^{ij}\p_iu_j\right] d^3x\,,
\]
where $u_\mu\equiv \p_\mu\theta+\alpha\p_\mu\beta$. Imposing the fluid domain to be $y\leq0$ and varying this action with respect to the fields $(\theta, \alpha, \beta, n, v^i, b_0)$ gives us
\begin{align}
\delta S_{\text{bulk}}&=\hbar\int dt\, dx\int\limits_{-\infty}^0dy\left\{n\left(\frac{m}{\hbar}v_i-u_i-\frac{\epsilon_{ij}}{2n}\p^jn\right)\delta v^i-\left(n-\frac{\nu e B}{2\pi\hbar}+\frac{\nu}{2\pi}\epsilon^{ij}\p_i u_j\right)\left(\delta b_0+ \p_t\beta\delta\alpha-\p_t\alpha\delta\beta\right)\right.     \nonumber
\\
&+\Big[\p_tn+\p_i(nv^i)\Big](\delta\theta+\alpha\delta\beta)+\frac{\nu}{2\pi}\left[\epsilon^{ij}\left(\p_ju_0-\p_tu_j+\p_jb_0-\frac{e}{\hbar}E_j\right)+\frac{2\pi}{\nu}nv^i\right](\p_i\alpha\delta\beta-\p_i\beta\delta\alpha) \nonumber
\\
&\left.-\left[u_0+b_0+v^iu_i-\frac{m}{2\hbar}v_i ^2+\frac{V'(n)}{\hbar}+\frac{\epsilon^{ij}}{2}\p_iv_j\right]\delta n\right\}-\hbar\int dt\, dx\left[\left(n v_y+\frac{\nu e}{2\pi\hbar} E_x\right)(\delta\theta+\alpha\delta\beta)\right. \nonumber
\\
&+\left.\frac{v_x}{2}\delta n+\frac{\nu}{2\pi}b_0(\p_x\alpha\,\delta\beta-\p_x\beta\,\delta\alpha)\right]\bigg|_{y=0}, \la{S-variation}
\end{align}
where we have used Faradays's law, i.e., $\p_tB+\epsilon^{ij}\p_iE_j=0$, and that $\p_ju_0-\p_tu_j=\p_j\alpha\p_t\beta-\p_j\beta\p_t\alpha$. 

Note that the bulk variation of $\theta$ gives us the continuity equation
\[
\p_tn+\p_i(nv^i)=0\,,
\]
whereas variation over $v^i$ provides us the velocity parametrization in terms of Clebsch potentials~(\ref{velocity}). The equation of motion for $b_0$ leads to the Hall constraint, which, in terms of the velocity field $v_i$, becomes
\[
n-\frac{eB}{2\pi\hbar}+\frac{\nu}{4\pi}\p_i\left(\frac{\p^in}{n}\right)+\frac{\nu m}{2\pi\hbar}\epsilon^{ij}\p_iv_j=0\,.
\] 
The Euler equation does not show up directly as an equation of motion of $S_{\text{bulk}}$, instead, it is obtained by combining all the other bulk equations of motion. Thus,
\begin{align}
\p_tv_i&=\frac{\hbar}{m}\left[\p_tu_i+\frac{\epsilon_{ij}}{2}\,\p^j\left(\frac{\p_t n}{n}\right)\right]=\frac{\hbar}{m}\left[\p_i(u_0+b_0)-\frac{e}{\hbar}E_i-\frac{2\pi}{\nu}\epsilon_{ij} n v^j-\frac{\epsilon_{ij}}{2}\,\p^j\left(\p_k v^k+\frac{v^k}{n}\p_kn\right)\right] \nonumber
\\
&=-\frac{1}{m}\p_i\left[V'+\frac{m}{2}v_j^2+\frac{\hbar}{2}\epsilon^{jk}\left(\p_jv_k-v_j\frac{\p_kn}{n}\right)\right]-\frac{e}{m}E_i-\frac{2\pi\hbar}{\nu m}\epsilon_{ij} nv^j-\frac{\hbar}{2mn}\p_j\left(n\epsilon_{ik}\p^k v^j\right)-\frac{\hbar \epsilon_{ij}}{2m}v^k\p^j\left(\frac{\p_k n}{n}\right).
\end{align}

After some algebra and using the identity
\be
\left(\epsilon_{jk}\p_i+\epsilon_{ij}\p_k+\epsilon_{ki}\p_j\right)\left(\frac{\p^j n}{n}\right)=0\,,
\ee
we finally obtain the Euler equation
\[
\p_tv_i=-v^j\p_j v_i-\frac{e}{m}\left(E_i+B\epsilon_{ij} v^j\right)-\frac{1}{mn}\p_i\left[n V'(n)-V(n)\right]+\frac{\hbar}{2mn}\p_j\left[n\left(\epsilon_{ik}\p^kv^j+\epsilon^{jk}\p_iv_k\right)\right].
\]

Let us now turn our attention to the boundary conditions. Since we are not fixing the field variation at the edge, the boundary conditions are obtained as the boundary equations of motion. For that, we also need to account for the edge action~(\ref{action-edge}), that is,
\[
S_{\text{edge}}=\frac{\hbar}{2}\int dt \, dx\,\left(\p_t\phi+\frac{\nu e A_0}{2\pi\hbar}\right)\left(\p_x\phi+\frac{ \nu e  A_x}{2\pi\hbar}-2\sqrt{n}\right)\bigg|_{y=0},
\]
and varying over $S_{\text{edge}}$ gives us
\be
\delta S_{\text{edge}}=-\hbar\int dt\, dx\left\{\frac{1}{2\sqrt{n}}\left(\p_t\phi+\frac{\nu e}{2\pi\hbar}A_0\right)\delta n-\left[\p_t\sqrt{n}-\p_x\left(\p_t\phi+\frac{\nu e}{2\pi\hbar}A_x\right)+\frac{\nu e}{4\pi\hbar}E_x\right]\delta\phi\right\}\bigg|_{y=0}.
\ee

Combining the edge terms of $\delta S_{\text{bulk}}+\delta S_{\text{edge}}$, we find that the boundary variation of $\theta$ provides us the anomaly inflow
\[
\left(n v_y+\frac{\nu e}{2\pi\hbar}E_x\right)\Big|_{y=0}=0\,,
\]
whereas the boundary variation of $n$ leads to the bosonization expression
\[
\left(\sqrt{ n}\, v_x+\p_t\phi+\frac{ \nu e}{2\pi\hbar} A_0\right)\Big|_{y=0}=0\,,
\]
in which the edge current is parametrized by the chiral boson field $\phi$. Combining it with the equation of motion for $\phi$, gives us the anomaly equation
\[
\left[\p_t(\sqrt{ n})+\p_x(\sqrt{ n}\, v_x)+\frac{\nu e}{4\pi\hbar} E_x\right]\Big|_{y=0}=0\,.
\]

%%%%%%%%%%%%%%%%%%%%%%%%%%%%%%
\section{Duality between CSGL theory and hydrodynamic action}
\label{app:duality}

In this section, we will work out the duality between the hydrodynamical action with topological terms and the Chern-Simon-Ginzburg-Landau theory for the Laughlin states. One more time, the fluid domain is taken to be $y\leq0$. Before proceeding, let us note that variation of $S_{\text{bulk}}$, i.e., Eq.~(\ref{S-variation}), provides us
\be
\p_x\alpha\p_t\beta-\p_t\alpha\p_x\beta+\p_xb_0-\frac{e}{\hbar}E_x-\frac{2\pi}{\nu}v_y=0\,.
\ee
Projecting it at the boundary and imposing the anomaly inflow condition, we end up with
\be
\left(\p_t\alpha \p_x\beta-\p_t\beta\p_x\alpha-\p_xb_0\right)\Big|_{y=0}=0\,. \la{BC-alpha-beta}
\ee
Therefore, we can rewrite $S_{\text{bulk}}$ as 
\be
S_{\text{bulk}}=S_{\text{hydro}}+S_{\text{top}}-\frac{\hbar\nu}{4\pi}\int dt\, dx\Big[\lambda\left(\p_t\alpha\p_x\beta-\p_x\alpha\p_t\beta-\p_xb_0\right)\Big]\bigg|_{y=0}, \la{S-bulk2}
\ee
since it reduces to $S_{\text{bulk}}= S_{\text{bulk}}+S_{\text{top}}$, after integrating $\lambda$ out. In fact, this is only possible because integrating out $\lambda$ imposes Eq.~(\ref{BC-alpha-beta}), which is compatible with the other boundary conditions.

Let us now turn our attention to $S_{\text{top}}$ and express it as
\begin{align}
S_{\text{top}}=&\,-\frac{\hbar\nu}{4\pi}\int\left[\left(b_0\,dt+\alpha d\beta-\frac{e}{\hbar}A\right)\wedge d\left(b_0\,dt+\alpha d\beta-\frac{e}{\hbar}A\right)\right]+\frac{\nu e^2}{4\pi\hbar}\int A\wedge dA \nonumber
\\
&-\frac{e\nu}{4\pi}\int dt\,dx\Big[(2\p_t\theta+b_0+\alpha\p_t\beta)A_x-(2\p_x\theta+\alpha\p_x\beta)A_0+b_0\,\alpha\p_x\beta\Big]\bigg|_{y=0}\,. 
\end{align}
Plugging this expression into Eq.~(\ref{S-bulk2}) and denoting $\theta=\vartheta+\lambda$, we end up with
\begin{align}
    S_{\text{bulk}}=&\,\,S_{\text{hydro}}-\frac{\hbar\nu}{4\pi}\int\left[\left(b_0\,dt+d\lambda+\alpha d\beta-\frac{e}{\hbar}A\right)\wedge d\left(b_0\,dt+d\lambda+\alpha d\beta-\frac{e}{\hbar}A\right)\right]+\frac{\nu e^2}{4\pi\hbar}\int A\wedge dA \nonumber
    \\
    &-\frac{e\nu}{4\pi}\int dt\,dx\Big[(2\p_t\vartheta+b_0+\p_t\lambda+\alpha\p_t\beta)A_x-(2\p_x\vartheta+\p_x\lambda+\alpha\p_x\beta)A_0+b_0\,\alpha\p_x\beta\Big]\bigg|_{y=0}\,. \la{S-bulk3}
\end{align}
Note that, with the exception of the very last term in the second line of Eq.~(\ref{S-bulk3}), $\lambda$, $b_0$, $\alpha$ and $\beta$ only appear in the combination $b_0 dt+d\lambda+\alpha d\beta$. The last term, however, vanish when imposing the boundary condition $b_0(x,0,t)=0$.

Although $b_0(x,0,t)=0$ is a particular of the boundary variation in Eq.~(\ref{S-variation}), it is not the unique solution. To ensure this boundary condition, we can either add a boundary term $\int dtdx\, \gamma\, b_0|_{y=0}$ in the hydrodynamic action or impose $b_0(x,0,t)=\delta b_0(x,0,t)=0$ by hand. In this section, we will consider the latter, even though the former provides the exactly same result. Hence, imposing $b_0(x,0,t)=0$ allow us to write the action $S_{\text{bulk}}$ in the form
\be
 S_{\text{bulk}}=S_{\text{hydro}}-\frac{\hbar\nu}{4\pi}\int_{\mathcal M}\left[a\wedge da-\frac{e^2}{\hbar^2}A\wedge dA+\zeta\wedge d\left(a-b_0 dt-\alpha d\beta+\frac{e}{\hbar}A\right)\right]-\frac{e\nu}{4\pi}\int_{\p\mathcal M} \Big[(2d\vartheta+a)\wedge A\Big],
\ee
where we denoted the fluid domain by $\mathcal M$ and $S_{\text{hydro}}$ is given by
\be
S_{\text{hydro}}=-\hbar\int_{\mathcal M}\left[n\left(\p_t\vartheta+a_0+\frac{e}{\hbar}A_0\right)+nv^i\left(\p_i\vartheta+a_i+\frac{e}{\hbar}A_i\right)-\frac{m}{2\hbar}nv_i^2+\frac{V(n)}{\hbar}+\frac{\epsilon^{ij}}{2}v_i\p_j n\right]d^3x\,.
\ee
Integrating $\zeta$ out gives us $a=b_0 dt+\alpha d\beta-\frac{e}{\hbar}A+d\lambda$, for some function $\lambda$, and we recover the action~(\ref{S-bulk3}) with the condition $b_0(x,0,t)=0$ imposed.

On the other hand, one could trace out the variables $b_0$, $\alpha$ and $\beta$. Integrating out $b_0$, imposing that $\delta b_0(x,0,t)=0$, gives us
\be
\epsilon^{ij}\p_i\zeta_j=0\,. \la{var-b0}
\ee
In addition to that, bulk variations of $\alpha$ and $\beta$ provide us
\begin{align}
&\epsilon^{ij}\left(\p_t\zeta_i-\p_i\zeta_0\right)\p_j\beta+\epsilon^{ij}\p_i\zeta_j\,\p_t\beta=0\,, \la{var-alpha}
\\
&\epsilon^{ij}\left(\p_t\zeta_i-\p_i\zeta_0\right)\p_j\alpha+\epsilon^{ij}\p_i\zeta_j\,\p_t\alpha=0\,, \la{var-beta}
\end{align}
respectively. Combining Eqs.~(\ref{var-b0}-\ref{var-beta}) and using that $\p_i\alpha$ and $\p_i\beta$ are linearly independent, we end up with 
\be
d\zeta=0\qquad \Longrightarrow\qquad \zeta=d\chi\,.
\ee

Therefore, after integrating out $b_0$, $\alpha$ and $\beta$, the action $S_{\text{bulk}}$ becomes
\be
S_{\text{bulk}}=S_{\text{hydro}}-\frac{\hbar\nu}{4\pi}\int_{\mathcal M}\left(a\wedge da-\frac{e^2}{\hbar^2}A\wedge dA\right)-\frac{\nu}{4\pi}\int_{\p\mathcal M} \Big[e\left(2d\vartheta+a-d\chi\right)\wedge A-\hbar \,d\chi\wedge a\Big].
\ee
Defining $\chi=\vartheta+\sigma$ and integrating out $\sigma$, we find that the topological hydrodynamical action $S_{\text{bulk}}$ is dual to the Chern-Simons action coupled to electromagnetic field $A_\mu$ and the matter fields $n$, $\vartheta$ and $v^i$, that is,
\be
S_{\text{bulk}}=S_{\text{hydro}}-\frac{\nu}{4\pi}\int\left[\hbar\left(a+d\vartheta\right)\wedge da-e\left(d\vartheta+\frac{e}{\hbar}A\right)\wedge dA\right]-\frac{\nu e}{4\pi}\int_{\p\mathcal M} a\wedge A\,. \la{S-bulk4}
\ee

To bring the action~(\ref{S-bulk4}) in the form of a CSGL action, we must trace out the velocity field and express the Madelung variables $n$ and $\vartheta$ in the form $\Phi=\sqrt{n}e^{i\vartheta}$. After some algebra, we find that
\begin{align}
S_{\text{bulk}}=&\,\int_{\mathcal M}\left[i\hbar\Phi^\dagger D_t\Phi-\frac{\hbar^2}{2m}|D_i\Phi|^2-V(|\Phi|)-\frac{|\Phi|^2}{m}\left(\hbar\epsilon^{ij}\p_ia_j+eB\right)\right]d^3x-\frac{i\hbar}{m}\int_{\p\mathcal M} d^2x\left(\Phi^\dagger  D_x\tilde\Phi\right)\Big|_{y=0}\nonumber
\\
&-\frac{\nu}{4\pi}\int_{\mathcal M}\left[\hbar\, a\wedge da-\frac{e^2}{\hbar}A\wedge dA-i\frac{\Phi^\dagger d\Phi-(d\Phi)^\dagger\Phi}{|\Phi|^2}\wedge\left(e\,dA-\hbar\,da\right)\right]-\frac{\nu e}{4\pi}\int_{\p\mathcal M} a\wedge A\,. \la{S-bulk5}
\end{align}
where $D_\mu\equiv\p_\mu+i\left(a_\mu+\tfrac{e}{\hbar}A_\mu\right)$ denotes the covariant derivative. Before proceeding to study $S_{\text{edge}}$, let us note that
\be
-\frac{i\nu e}{4\pi}\int_{\mathcal M}\frac{\Phi^\dagger d\Phi-(d\Phi)^\dagger\Phi}{|\Phi|^2}\wedge\,dA\approx-\frac{i\hbar}{2}\int_{\mathcal M}\left(\Phi^\dagger \p_i\Phi-\p_i\Phi^\dagger\,\Phi\right)\epsilon^{ij}\frac{E_j}{B}\,d^3x\,, \la{NC}
\ee
where we have used the Hall constraint to approximate $|\Phi|^2\approx \frac{\nu e B}{2\pi\hbar}$. A similar term as the one in the RHS of Eq.~(\ref{NC}) has appeared before in the context of composite fermions~\cite{nguyen2018particle}, where the authors used the Newton-Cartan theory to introduce it. Such term is in fact fundamental to get the Girvin-MacDonald-Platzman (GMP) algebra for the projected density.

Let us now focus on $S_{\text{edge}}$ and express it in terms of $\Phi$. Thus, 
\be
S_{\text{edge}}=\frac{\hbar}{2}\int_{\mathcal M} d^2x\,\left(\p_t\phi+\frac{\nu e A_0}{2\pi\hbar}\right)\left(\p_x\phi+\frac{ \nu e  A_x}{2\pi\hbar}-2|\Phi|\right)\bigg|_{y=0}, \la{S-edge2}
\ee
and the flux attachment action arising from the duality with the topological fluid action becomes simply $S=S_{\text{bulk}}+S_{\text{edge}}$, where $S_{\text{bulk}}$ and $S_{\text{edge}}$ are given by Eqs.~(\ref{S-bulk5}, \ref{S-edge2}) respectively.

It is worth noting that the coupling between the chiral boson $\phi$ and the condensate field $\Phi$ is somewhat unusual from the chiral Luttinger liquid point of view, but it arises naturally in the hydrodynamic framework. Also, differently from the TQFT phenomenology, in the CSGL theory derived from the topological fluid action, the statistical field $a_\mu$ do not couple directly to the chiral boson at the edge.

%%%%%%%%%%%%%%%%%%%%%%%%
%%%%%%%%%%%%%%%%%%%%%%%%
%\begin{thebibliography}{1}

%\bibitem{nguyen2018particle} Nguyen, D. X., Golkar, S., Roberts, M. M. and Son, D. T., Phys. Rev. B {\bf 97}, 195314 (2018).

%\bibitem{nguyen2018particle} Nguyen, D. X., Golkar, S., Roberts, M. M. and Son, D. T., ``\emph{Particle-hole symmetry and composite fermions in fractional quantum Hall states}", Phys. Rev. B {\bf 97}, 195314 (2018).

%\end{thebibliography} 

\twocolumngrid

\bibliography{oddviscosity-bibliography.bib}
%%%%%%%%%%%%%%%%%%%%%%%%
%%%%%%%%%%%%%%%%%%%%%%%%

\end{document}